# Frequency-domain formulation of signal propagation in multistrip Resistive Plate Chambers and its low-loss, weak-coupling analytical approximation

**P. Fonte**

*LIP-Laboratório de Instrumentação e Física Experimental de Partículas*
*Dep. de Física, Univ. de Coimbra, 3004-516 Coimbra, Portugal*
*ISEC-Instituto Superior de Engenharia de Coimbra*
*Rua Pedro Nunes - Quinta da Nora, 3030-199 Coimbra, Portugal*
*E-mail:* `fonte@coimbra.lip.pt`

In this study the signal propagation in multistrip Resistive Plate Chambers is formulated exactly in the framework of the multiconductor transmission line theory in the frequency domain, allowing losses to be incorporated. For the case of weak-coupling and low-losses a first-order expansion yields simple, fully analytical, expressions that include reflections. In this approximation the modal spectrum can be calculated analytically as well, allowing to estimate easily the strength of the modal dispersion phenomenon.

## 1. Introduction

In Resistive Plate Chambers (RPCs), the charge induction process injects a current into the readout electrodes, which must then be transmitted towards the amplifiers. This is particularly important for timing RPCs, where the use of a narrow gas gap generates a very fast signal upon which a precise time is eventually measured. Often these electrodes take the form of strip arrays for which a well developed theory of Multiconductor Transmission Lines (MTL) is available (e.g. [1], [2]). The theory of MTLs has been particularized for the case of RPCs in [3] and further developed in [4].

However, exact analytical solutions of reasonable transparency cannot be derived in the general case and only numerical evaluation is feasible. One possible strategy is to consider exact solutions in situations sufficiently simplified to yield useful analytical expressions, as done in [4], where it is treated the case of 5 strips with nearest-neighbour coupling only. In here, for the same purpose, we will opt to linearize the exact frequency-domain solutions (explicitly including losses) and derive systematically first-order approximations, yielding simple analytical expressions for the signals, modal spectrum and crosstalk. These approximations are checked against the exact solution in a number of relevant examples.

It is worth mentioning that the standard MTL theory is likely inaccurate for widespread structures such as RPCs, as it misses the fact that the strips are quite separated in space and that there must be some propagation velocity across the strips as well, which is absent from the MTL theory. This may limit the formation of fully organized propagation modes involving all strips of a wide array. Apart from an exact theory, an MTL nearest-neighbours formulation, as discussed in [3], [4] and in the present work, may turn out to be more satisfactory, but further work is needed in this direction.

## 2. General theory

The system consists of $N+1$ translationally invariant electrodes (array of strips), including one reference ("ground") electrode, spanning in length from $x=0$ to $x=D$. The currents in the strips and the strip-to-ground and termination-to-ground voltages are denoted by the $N \times 1$ vectors $\mathbf{I}(x,t)$, $\mathbf{V}(x,t)$ and $\mathbf{V}_{T \in \{0,D\}}(t)$. A current pulse $I_0(t)$ is injected from the ground into a single "driven" strip at position $x_0$ (or more, by superposition), propagating in both directions along $x$ and coupling to the neighbouring strips (crosstalk). The strips are terminated by resistor networks (or just the input impedance of the amplifiers) at $x=0$ and $x=D$. These terminations don't need to be equal and can be extended to handle general linear networks and interconnections with other MTLs [5].

The main variables of the problem are represented in Fig. 1.

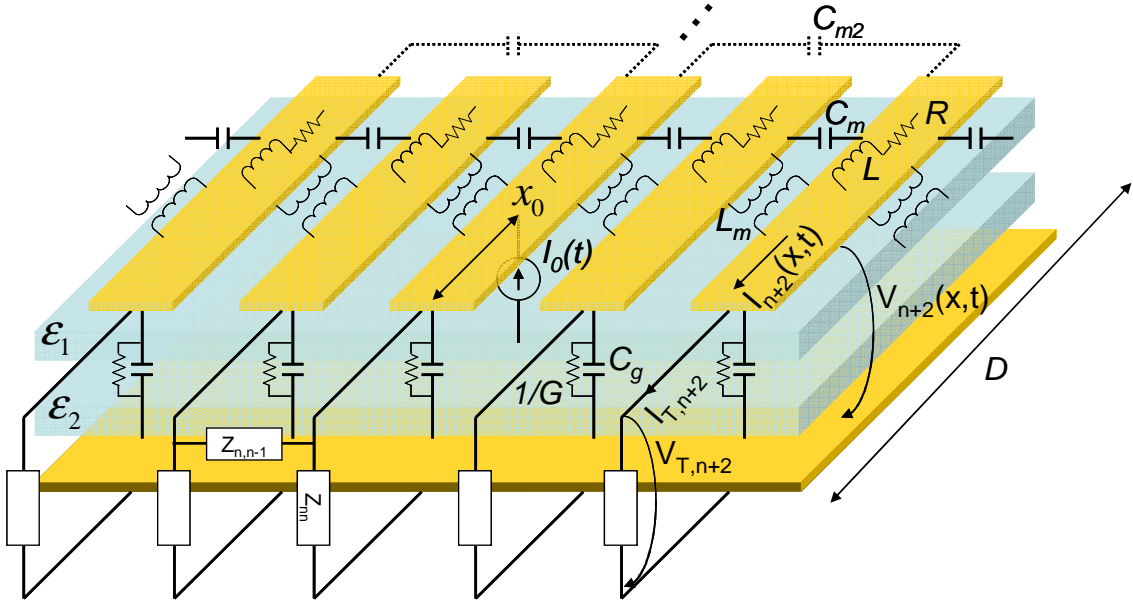

Fig. 1 – Representation of the geometry and the most relevant quantities for the problem of propagation in a strip array.

The line electrical properties are defined by the $N \times N$ per unit length capacitance, inductance, resistance and conductance matrices, respectively $\mathbf{C}, \mathbf{L}, \mathbf{R}, \mathbf{G}$. In general, these are symmetric matrices and all conductors will couple to each other via the diagonal elements, but later we will particularize for a "weak-coupling" approximation.

If all materials involved have magnetic permeability close to vacuum's, the inductance matrix can be calculated from the capacitance matrix $\mathbf{C}_0$ obtained with all dielectrics removed

$$\mathbf{L} = \mathbf{C}_0^{-1} / c^2, \tag{1}$$

$c$ being the speed of light in the vacuum. Therefore, all parameters can be obtained by electrostatic[*] calculations only.

The fundamental "telegrapher's" equations to be satisfied are

$$\frac{\partial^2 \mathbf{V}}{\partial x^2} - \mathbf{LC}\frac{\partial^2 \mathbf{V}}{\partial t^2} = \mathbf{RGV} + (\mathbf{RC} + \mathbf{LG})\frac{\partial \mathbf{V}}{\partial t}$$
$$\frac{\partial^2 \mathbf{I}}{\partial x^2} - \mathbf{CL}\frac{\partial^2 \mathbf{I}}{\partial t^2} = \mathbf{GRI} + (\mathbf{CR} + \mathbf{GL})\frac{\partial \mathbf{I}}{\partial t} \tag{2}$$

In the time domain these can be solved analytically only in the lossless case, vanishing the right hand side (rhs), while in the frequency domain losses can be accounted for. Following closely [2], one starts by considering the *ansatz*

$$\mathbf{V}_m^\pm(x,t) = \int_{-\infty}^{\infty} \mathbf{V}_m^\pm(x,\omega) e^{i\omega t} d\omega$$
$$\mathbf{I}_m^\pm(x,t) = \int_{-\infty}^{\infty} \mathbf{I}_m^\pm(x,\omega) e^{i\omega t} d\omega \tag{3}$$
$$\mathbf{V}_m^\pm(x,\omega) = F(\omega) e^{\mp \gamma_m x} \hat{\mathbf{V}}_m(\omega) K_{V,m}$$
$$\mathbf{I}_m^\pm(x,\omega) = \pm F(\omega) e^{\mp \gamma_m x} \hat{\mathbf{I}}_m(\omega) K_{I,m}$$

where $F(\omega)$ is the Fourier transform of the line excitation signal $I_0(t)$ and $i$ the imaginary unit. The sign conventions mean that in general there will be signals propagating in both directions ($\pm x$) along the line and that the currents propagating to the left ($x < x_0$, - wave) have a sign opposite to that of the corresponding voltage. The index $m$ denotes the "propagation mode" and $K_{\{I,V\},m}$ denotes the respective strength. This leads to the interrelated *eigenvalue* problems

$$\gamma_m^2 \hat{\mathbf{V}}_m(\omega) = \mathbf{ZY}\hat{\mathbf{V}}_m(\omega)$$
$$\gamma_m^2 \hat{\mathbf{I}}_m(\omega) = \mathbf{YZ}\hat{\mathbf{I}}_m(\omega) \tag{4}$$

involving the line impedance and admittance matrices

---

[*] The calculations needed to calculate $\mathbf{R}$ and $\mathbf{G}$ are essentially the same as for electrostatics. See [8] sect. 1.3, 2.7. and [4].

$$\begin{aligned}\mathbf{Z} &= \mathbf{R} + i\omega\mathbf{L} \\ \mathbf{Y} &= \mathbf{G} + i\omega\mathbf{C}\end{aligned} \quad (5)$$

on which all quantities can be frequency-dependent. Indeed, from this point onwards, almost all quantities are a function of $\omega$ so this will not be explicitly indicated.

The propagation coefficients $\gamma_m$ combine the signal attenuation coefficient and the propagation velocity in their real and imaginary parts:

$$\gamma_m = \alpha_m + i\beta_m = \alpha_m + i\frac{\omega}{v_m} \quad (6)$$

where $v_m$ is the modal phase velocity.

The voltage and current modal (column) vectors $\hat{\mathbf{V}}_m, \hat{\mathbf{I}}_m$, the solutions of (4), represent special combinations of line voltages and currents ("modes") that propagate with the same propagation coefficient $\gamma_m$. Except in degenerate cases, for a system with $N$ strips there will be $N$ propagation modes. The matrices formed by the columnwise concatenation of the modal vectors ($\hat{\mathbf{V}}, \hat{\mathbf{I}}$) form a basis of the space of all possible line excitations. The actual excitation determines the relative strength $K_m$ of the modes that will carry the signal along the line and the observable signal will be the sum of all modes':

$$\begin{aligned}\mathbf{V}(x) &= \sum_{m=1}^{N} \mathbf{V}_m^+(x) + \mathbf{V}_m^-(x) \\ \mathbf{I}(x) &= \sum_{m=1}^{N} \mathbf{I}_m^+(x) + \mathbf{I}_m^-(x)\end{aligned} \quad (7)$$

Note that each mode propagates with a different velocity, leading, after some propagation distance, to a decoherence effect called modal dispersion.

Voltages and currents in the line are related by the characteristic impedance $\mathbf{Z}_c$ as $\mathbf{V}^\pm(x,t) = \pm \mathbf{Z}_c \mathbf{I}^\pm(x,t)$ (therefore $\hat{\mathbf{V}} = \mathbf{Z}_c \hat{\mathbf{I}}$). This is given by

$$\mathbf{Z}_c = \hat{\mathbf{V}}\boldsymbol{\gamma}^{-1}\hat{\mathbf{V}}^{-1}\mathbf{Z} = \mathbf{Y}^{-1}\hat{\mathbf{I}}\boldsymbol{\gamma}\hat{\mathbf{I}}^{-1} \quad (8)$$

where $\boldsymbol{\gamma}$ is the diagonal matrix with entries $\gamma_m$.

In RPCs the modal excitation is determined by the boundary condition that the current $I_0(t)$ is injected uniquely in the $n_{th}$ strip, propagating equally in both directions: $\mathbf{I}^+(x_0,t) - \mathbf{I}^-(x_0,t) = [0,\cdots,I_0(t),\cdots,0]^T$ or, in the frequency domain, $\mathbf{I}^+(x_0) - \mathbf{I}^-(x_0) = [0,\cdots,F,\cdots,0]^T$. Consequently, the initial line excitation will be

$$\mathbf{I}_m^\pm(x) = \pm e^{\mp \gamma_m (x-x_0)} \frac{F}{2} \hat{\mathbf{I}}_m \hat{I}_{m,n}^{-1}$$
$$\mathbf{V}_m^\pm(x) = e^{\mp \gamma_m (x-x_0)} \frac{F}{2} \mathbf{Z}_c \hat{\mathbf{I}}_m \hat{I}_{m,n}^{-1} \qquad (9)$$

applying the + index to $x > x_0$ and vice-versa. It can be seen that the current modal strengths are given by the rows $m$ of the $n_{th}$ column of the inverse of the current modal matrix, $K_{I,m} = \hat{I}_{m,n}^{-1}$, causing the modes to interfere destructively at $x = x_0$ for all strips except the $n_{th}$. Modal dispersion will cause decoherence along the line, eventually destroying this cancelation. The signals loose their initial shape because the modes contributing to their propagation decohere and crosstalk increases by lack of cancellation in the non-excited strips (see for instance, in [4] fig. 8, [3] fig. 7).

Both waves will eventually reach the end of the line and a part of the incident current will be reflected and its complement will flow on the respective terminations. These may be described by their Z-representation $\mathbf{Z}_{T \in \{0,D\}}$, applying:

$$\begin{aligned}\mathbf{V}_{inc} + \mathbf{V}_{ref} &= \mathbf{V}_T = \mathbf{Z}_T \mathbf{I}_T = \mathbf{Z}_T \left( \mathbf{I}_{inc} + \mathbf{I}_{ref} \right) \\ \mathbf{V}_{inc} &= \mathbf{Z}_c \mathbf{I}_{inc}, \mathbf{V}_{ref} = -\mathbf{Z}_c \mathbf{I}_{ref}\end{aligned} \qquad (10)$$

This leads to the voltage and current reflection and transmission matrices $\mathbf{\Gamma}_{V,I}, \mathbf{T}_{V,I}$

$$\begin{aligned}\mathbf{V}_{ref} &= \mathbf{\Gamma}_V \mathbf{V}_{inc} = \left[ (\mathbf{Z}_T - \mathbf{Z}_c)(\mathbf{Z}_T + \mathbf{Z}_c)^{-1} \right] \mathbf{V}_{inc} \\ \mathbf{I}_{ref} &= -\mathbf{\Gamma}_I \mathbf{I}_{inc} = \left[ (\mathbf{Z}_T + \mathbf{Z}_c)^{-1} (\mathbf{Z}_T - \mathbf{Z}_c) \right] \mathbf{I}_{inc} \\ \mathbf{V}_T &= \mathbf{T}_V \mathbf{V}_{inc} = (\mathbf{1} + \mathbf{\Gamma}_V) \mathbf{V}_{inc} = \left[ 2\mathbf{Z}_T (\mathbf{Z}_T + \mathbf{Z}_c)^{-1} \right] \mathbf{V}_{inc} \\ \mathbf{I}_T &= \mathbf{T}_I \mathbf{I}_{inc} = (\mathbf{1} - \mathbf{\Gamma}_I) \mathbf{I}_{inc} = \left[ 2(\mathbf{Z}_T + \mathbf{Z}_c)^{-1} \mathbf{Z}_c \right] \mathbf{I}_{inc}\end{aligned} \qquad (11)$$

Note that if $\mathbf{Z}_c$ commutes with $\mathbf{Z}_T$ the reflection matrices for current and voltage are identical. Important cases are $\mathbf{Z}_T = \mathbf{Z}_c$, the perfect termination for which the reflection coefficients are null, and $\mathbf{Z}_T = \mathbf{1} R_T$ ($\mathbf{1}$ = unit $N \times N$ matrix): termination by identical single resistors to ground.

Lets be interested in the signal present in the termination at $x = D$ (D-end). The + wave in (9) reaches directly this end, while the - wave will be reflected at $x = 0$ (0-end) and will equally reach the D-end. For each mode, the current incident on the termination after at most a single reflection in the 0-end will be

$$\mathbf{I}_m^*(D) = \underbrace{e^{-\gamma_m(D-x_0)}\frac{F}{2}\hat{\mathbf{I}}_m\hat{I}_{m,n}^{-1}}_{+\text{wave propagates to the D-end}} + \underbrace{e^{-\gamma_m D}}_{+\text{wave propagates back}}\underbrace{(-\mathbf{\Gamma}_{I,0})}_{\text{reflection}}\underbrace{\left(-e^{-\gamma_m x_0}\frac{F}{2}\hat{\mathbf{I}}_m\hat{I}_{m,n}^{-1}\right)}_{-\text{wave propagates to the 0-end}}$$

$$= \frac{F}{2}\left(\mathbf{1}e^{-\gamma_m(D-x_0)} + \mathbf{\Gamma}_{I,0}e^{-\gamma_m(D+x_0)}\right)\hat{\mathbf{I}}_m\hat{I}_{m,n}^{-1} \qquad (12)$$

Remember that phase shifts in the frequency domain correspond to delays in the time domain. This incident wave will be now reflected in the D-end and again in the 0-end, reaching again the D-end, etc. So, the total wave incident in the D-end for each mode, including all reflections is given by a geometric series:

$$\mathbf{I}_m(D) = \left[\sum_{j=0}^{\infty}\left(\mathbf{\Gamma}_{I,0}\mathbf{\Gamma}_{I,D}e^{-\gamma_m 2D}\right)^j\right]\mathbf{I}_m^*(D) = \left(\mathbf{1} - \mathbf{\Gamma}_{I,0}\mathbf{\Gamma}_{I,D}e^{-\gamma_m 2D}\right)^{-1}\mathbf{I}_m^*(D) \qquad (13)$$

Finally, the (Fourier transform of the) current $\mathbf{I}_D$ and of the voltage $\mathbf{V}_D$ in the D-termination will be

$$\mathbf{Z}_D^{-1}\mathbf{V}_D = \mathbf{I}_D = \mathbf{T}_{I,D}\mathbf{I}(D) = \mathbf{T}_{I,D}\sum_{m=1}^{N}\mathbf{I}_m(D)$$
$$= F\frac{\mathbf{T}_{I,D}}{2}\sum_{m=1}^{N}\left(\mathbf{1} - \mathbf{\Gamma}_{I,0}\mathbf{\Gamma}_{I,D}e^{-\gamma_m 2D}\right)^{-1}\left(\mathbf{1}e^{-\gamma_m(D-x_0)} + \mathbf{\Gamma}_{I,0}e^{-\gamma_m(D+x_0)}\right)\hat{\mathbf{I}}_m\hat{I}_{m,n}^{-1} \qquad (14)$$

which is our main result: the frequency-domain counterpart of eq. (13) of [4], calculated for the D-end instead, generalized for arbitrary terminations and summed over all reflections. Of course, the terms that multiply $F$ express the frequency response of the line to the excitation current, that is, its transfer function.

The same result can be expressed in terms of voltage waves

$$\mathbf{V}_D = F\frac{\mathbf{T}_{V,D}}{2}\sum_{m=1}^{N}\left(\mathbf{1} - \mathbf{\Gamma}_{V,0}\mathbf{\Gamma}_{V,D}e^{-\gamma_m 2D}\right)^{-1}\left(\mathbf{1}e^{-\gamma_m(D-x_0)} + \mathbf{\Gamma}_{V,0}e^{-\gamma_m(D+x_0)}\right)\mathbf{Z}_c\hat{\mathbf{I}}_m\hat{I}_{m,n}^{-1} \qquad (15)$$

which corresponds to eq. (10) of [4], calculated for the D-end and summed over all reflections.

It is clear that the general solution (14) is capable of describing a very rich behaviour arising from the combination of many different parameters. Exact analytical solutions of reasonable transparency cannot be derived, largely because of the eigenvector calculation (4). In the next section we linearize the exact solutions and derive systematic first-order approximations. For this it is useful to take a step back and concentrate on the discrete reflections from each initial ± wave. From (13) we may write for the $j_{th}$ reflection of each wave

$$\mathbf{I}_D^+(j) = F \frac{\mathbf{T}_{I,D}}{2} \left( \mathbf{\Gamma}_{I,0} \mathbf{\Gamma}_{I,D} \right)^j \hat{\mathbf{I}} \mathbf{U}^+ \hat{\mathbf{I}}_n^{-1}$$

$$\mathbf{I}_D^-(j) = F \frac{\mathbf{T}_{I,D}}{2} \left( \mathbf{\Gamma}_{I,0} \mathbf{\Gamma}_{I,D} \right)^j \mathbf{\Gamma}_{I,0} \hat{\mathbf{I}} \mathbf{U}^- \hat{\mathbf{I}}_n^{-1} \qquad (16)$$

$$\mathbf{U}^\pm(j) = \operatorname{diag}\left( e^{-\gamma_m (2jD + D \mp x0)} \right)$$

where $\mathbf{U}^\pm$ is a diagonal matrix of phases. Concerning the placement of $\mathbf{U}^\pm$, see [4] eq. (10). In case the reflection and transmission matrices are identical on both ends this can be written more compactly:

$$\mathbf{I}_D^+(j) = F \frac{\mathbf{H}(2j)}{2} \hat{\mathbf{I}} \mathbf{U}^+ \hat{\mathbf{I}}_n^{-1}$$

$$\mathbf{I}_D^-(j) = F \frac{\mathbf{H}(2j+1)}{2} \hat{\mathbf{I}} \mathbf{U}^- \hat{\mathbf{I}}_n^{-1} \qquad (17)$$

$$\mathbf{H}(k) = \mathbf{T}_I \mathbf{\Gamma}^k$$

When the modal behaviour can be neglected (considering all $\gamma_m$ identical, $\gamma_m = \gamma$) then the $\mathbf{U}^\pm$ are just global phases, corresponding to a single time delay and attenuation for each reflection. In this case $\hat{\mathbf{I}} \mathbf{U}^\pm \hat{\mathbf{I}}_n^{-1} = \left[ 0, ..., e^{-\gamma(2jD + D \mp x0)}, ..., 0 \right]$, so the effect of the line on the injected signal is determined essentially by the $n_{th}$ column of $\mathbf{H}(k)$, applying the even $k$ to the + wave and its reflections and the odd $k$ to the - wave and its reflections.

Naturally the full signal will be the sum of all reflections:

$$\mathbf{I}_D = \sum_{j=0}^{\infty} \mathbf{I}_D^+(j) + \mathbf{I}_D^-(j). \qquad (18)$$

**2.1. Weak-coupling low-loss approximation**

In a planar repetitive structure such as depicted in Fig. 1 it is clear that the direct coupling from a strip to its second neighbour, represented by $c_{m2}$, will be much smaller than to its nearest neighbour and may be neglected. For instance, the values given in [4] table 2 indicate that $C_m / C \approx 1/10$, $C_{m2}/C_m \ll 1/10$. Furthermore, realistic numerical examples [3], [4], suggest that the nearest-neighbour coupling strength is relatively weak, on the order of 10%. This suggests a number of simplifications that yield analytical solutions.

The minimal weak-coupling approximation would be to consider only interactions between nearest neighbours, neglecting $c_{m2}$. In this approximation the capacity matrix becomes a symmetric tridiagonal Toeplitz (STT) matrix:

$$\mathbf{C} = \begin{bmatrix} C & -C_m & 0 & 0 & \\ -C_m & C & -C_m & 0 & \cdots \\ 0 & -C_m & C & -C_m & \\ 0 & 0 & -C_m & C & \\ & \vdots & & & \ddots \end{bmatrix} \equiv \{C, \ -C_m\} = C\{1, \ -u\} \quad (19)$$

$$C = C_g + C_m, \quad u = C_m / C$$

As most of this section will deal with such type of matrices these will be represented in curly braces as exemplified above. The results will not depend on the size of the matrix and the $2 \times 2$ case lossless case corresponds to the exact solution of a lossless 2-strip system, reproducing eqs. (6), (7) and (11) in [6].

Besides $\mathbf{C}$ we need the capacitance matrix with all dielectrics removed

$$\mathbf{C}_0 = \{C_0, \ -C_{0m}\} = C_0\{1, \ -u_0\} \quad (20)$$

and the loss matrices, to be considered as vanishingly small (weak losses)

$$\mathbf{R} = \{r, \ 0\}, \ \mathbf{G} = \{g, \ 0\} \quad (21)$$

and single-resistor terminations to ground (so, all reflection matrices will be identical)

$$\mathbf{Z}_T = \{R_T, \ 0\}. \quad (22)$$

Requiring additionally $u, u_0 \ll 1$ (weak coupling), all expressions will be developed in series of $u, u_0, r, g$ and truncated to the first order. This renders all matrices STT, as the elements further removed from the main diagonal are at least quadratic in these variables.

Under these approximations

$$\mathbf{L} = \{L, \ L_m\} = \frac{1}{c^2 C_0}\{1, \ u_0\} \quad (23)$$

$$\mathbf{YZ} = \beta\{2i\alpha - \beta, \ \beta(u - u_0)\}, \quad \alpha = \frac{rC + gL}{\sqrt{LC}}, \beta = \omega\sqrt{LC}. \quad (24)$$

The squared propagation coefficients and the modal currents are the eigenvalues and the eigenvectors of (4). For an STT matrix $\mathbf{YZ} = \{a, \ b\}$ of size $N$ there are simple analytical expressions for these [7]:

$$\gamma_m^2 = a + 2b\cos\left(\frac{m\pi}{N+1}\right), \ \hat{I}_{k,m} = \sqrt{\frac{2}{N+1}}\sin\left(\frac{km\pi}{N+1}\right), \ m,k = 1,2,...,N \quad (25)$$

Furthermore, $\hat{\mathbf{I}}$ is unitary and symmetric, so $\hat{\mathbf{I}}^{-1} = \hat{\mathbf{I}}$. For instance, a $5 \times 5$ system yields

$$\gamma^2 = a + b \begin{bmatrix} \sqrt{3} \\ 1 \\ 0 \\ -1 \\ -\sqrt{3} \end{bmatrix}, \quad \hat{\mathbf{I}} = \begin{bmatrix} \dfrac{\sqrt{3}}{6} & \dfrac{1}{2} & \dfrac{\sqrt{3}}{3} & \dfrac{1}{2} & \dfrac{\sqrt{3}}{6} \\ \dfrac{1}{2} & \dfrac{1}{2} & 0 & -\dfrac{1}{2} & -\dfrac{1}{2} \\ \dfrac{\sqrt{3}}{3} & 0 & -\dfrac{\sqrt{3}}{3} & 0 & \dfrac{\sqrt{3}}{3} \\ \dfrac{1}{2} & -\dfrac{1}{2} & 0 & \dfrac{1}{2} & -\dfrac{1}{2} \\ \dfrac{\sqrt{3}}{6} & -\dfrac{1}{2} & \dfrac{\sqrt{3}}{3} & -\dfrac{1}{2} & \dfrac{\sqrt{3}}{6} \end{bmatrix} \qquad (26)$$

Because $\hat{\mathbf{I}}^{-1} = \hat{\mathbf{I}}$, the columns represent the current ratios for each mode and simultaneously the amount of excitation of each mode when the strip corresponding to a column is driven. If this is the central strip it can be seen that all odd modes are excited[*] with equal amplitudes and the central mode inverted. It is also clear that within these approximations $\hat{\mathbf{I}} = \hat{\mathbf{V}}$.

Taking into consideration (24) and (25), the propagation coefficients are given by (always to the first-order)

$$\gamma_m = \alpha + i\beta \left[ 1 - (u - u_0) \cos\left(\frac{m\pi}{N+1}\right) \right] \qquad (27)$$

and therefore, comparing with (6), the modal velocity spectrum is contained within

$$\frac{\overline{v} - \Delta v}{\overline{v}} = 1 - |u_0 - u| < \frac{v_m}{\overline{v}} < 1 + |u_0 - u| = \frac{\overline{v} + \Delta v}{\overline{v}}$$
$$\overline{v} = \frac{1}{\sqrt{LC}}, \qquad \Delta v = \overline{v}|u_0 - u| \qquad (28)$$

and it is not affected by the losses, represented by $\alpha$. If $u_0 = u$, indicating a dielectrically homogeneous or compensated ([6], [8]) system, the spectrum collapses to a single velocity and there are no modes. In principle, even in this case there is some signal dispersion $v(\omega)$ related to the $r, g$ parameters, but this is null in first order.

Within the approximations, the structure of the characteristic impedance matrix $\mathbf{Z}_c$ doesn't depend on the details of the modal structure and it can be expressed as

$$\mathbf{Z}_c = \underbrace{\sqrt{\frac{L}{C}}}_{\overline{Z}_c} \left\{ 1 - i\frac{\delta}{2\beta}, \ \frac{u + u_0}{2} \right\}, \ \delta = \frac{rC - gL}{\sqrt{LC}}, \qquad (29)$$

---

[*] All even modes are zero for the center strip, so can't contribute. The modes are clearly related to the Fourier series.

Note that the imaginary, capacitive-like, term can be compensated by making $rC = gL$ [*].

The current reflection and transmission coefficients are

$$\mathbf{\Gamma} = \left\{ \frac{1-w}{1+w} + i\frac{w}{(1+w)^2}\frac{\delta}{\beta},\quad -\frac{w}{(1+w)^2}(u+u_0) \right\},\ w = \frac{Z_c}{R_T} \tag{30}$$

$$\mathbf{1} - \mathbf{\Gamma} = \mathbf{T}_I = \frac{w}{1+w}\left\{ 2 - i\frac{1}{1+w}\frac{\delta}{\beta},\quad \frac{1}{1+w}(u+u_0) \right\} \tag{31}$$

and the important matrix $\mathbf{H}(k)$ (see (17)) is

$$\mathbf{H}(k) = \{H,\ H_m\} = \frac{w(1-w)^k}{(1+w)^{k+1}}\left\{ 2 - i\frac{1-(2k+1)w}{1-w^2}\frac{\delta}{\beta},\quad \frac{1-(2k+1)w}{1-w^2}(u+u_0) \right\} \tag{32}$$

Besides the present first-neighbour calculation, one may consider a generalization for all strips in the form of successive nearest-neighbour couplings

$$\mathbf{H}_{all}(j) \approx \begin{bmatrix} H & H_m & H_m\eta & H_m\eta^2 & \cdots \\ H_m & H & H_m & H_m\eta & \\ H_m\eta & H_m & H & H_m & \\ H_m\eta^2 & H_m\eta & H_m & H & \\ \vdots & & & & \ddots \end{bmatrix}, \tag{33}$$

on which the tridiagonal part still corresponds to $\mathbf{H}(j)$ and $\eta$ is the crosstalk ratio (the imaginary, capacitive part cancels to the first order)

$$\eta(k) = \frac{H_m}{H} = \frac{1-(2k+1)w}{1-w^2}\frac{u+u_0}{2} \tag{34}$$

Note that the relative crosstalk tends to increase with successive reflections indicated by $k$, but, if $w < 1$, it may decrease for small $k$ and change sign, increasing afterwards.

## 3. Discussion

The general properties of the transmission line transfer function are well known (e.g. [4], [8]) and are not particularly illuminating for broadband impulse signals. In this discussion we will concentrate on the time-domain, as this is of straightforward interpretation.

For the purpose of illustrating the features of the exact and approximate solutions found, several examples were calculated using the parameters listed in Table 1. The values of some important derived quantities are presented as well.

---

[*] The historical Heaviside condition for undistorted signal transmission, but this applies mostly to $r, g$-related dispersion, which is zero in first order, so it is neglected here.

Table 1 - Parameters for the examples presented in the discussion. The $\mathbf{C}$ values for sets $\mathcal{A}$ and $\mathcal{B}$ were taken from Table 2 in [4] (uncompensated micro-stripline) and the $\mathbf{C}_0$ values were mathematically adjusted for exact compensation while keeping the same $\overline{v}$. The $\mathbf{C}$ and $\mathbf{C}_0$ values for sets $\mathcal{C}$ and $\mathcal{D}$ were taken from the same table, respectively uncompensated micro-stripline and compensated stripline. The $r$ value was chosen just to yield a small, but not negligible, value for $\alpha$ and $\delta$. The value of $R_T$ was chosen to almost cancel the $k=2$ reflection in sets $\mathcal{A}$-$\mathcal{C}$, highlighting this feature of the solutions.

|  | $\mathcal{A}$ | $\mathcal{B}$ | $\mathcal{C}$ | $\mathcal{D}$ |
|---|---|---|---|---|
|  | Basic parameters ||||
| $C$ (pF/m) | 276 ||| 390 |
| $C_m$ (pF/m) | 36.1 ||| 28.3 |
| $C_{m2}$ (pF/m) | 0 | 1.23 || 0.0004 |
| $C_0$ (pF/m) | 98.7 ||| 131 |
| $C_{0m}$ (pF/m) | 12.9 || 15.7 | 9.50 |
| $C_{0m2}$ (pF/m) | 0 | 0.440 | 1.21 | 0.0003 |
| $r$ (Ω/m) | 0 | 1 |||
| $g$ (S/m) | 0 ||||
| $R_T$ (Ω) | 100 ||||
| $D$ (m) | 2 ||||
| $x_0$ (m) | 0.8 ||||
| $\sigma$ (ns) | 1.25 ||||
|  | Derived quantities ||||
| $\alpha$ (m$^{-1}$) | 0 | 0.050 || 0.034 |
| $\delta$ (m$^{-1}$) | 0 | 0.050 || 0.034 |
| $\overline{v}$ (cm/ns) | 17.9 ||| 17.4 |
|  |  |  |  |  |
| $w$ | 0.20 ||| 0.15 |
| $u-u_0$ | 0 || -0.0485 | 45 x10$^{-6}$ |
| $u+u_0$ | 0 | 0.26 | 0.31 | 0.15 |
| $\Delta t$ (ns) | 0 | 0 | 0.32 | 3.1x10$^{-3}$ |
| $\Theta(1,2,3)$ | 0 | 0 | 0.68, 2.8, 11 | 0.012, 0.090, 0.79 |

For clarity, it was used in the examples an idealized "mathematical signal", similar to a real one, as shown in Fig. 2. The idealized signal has a half-width at half-maximum $\sigma = 1.25$ ns.

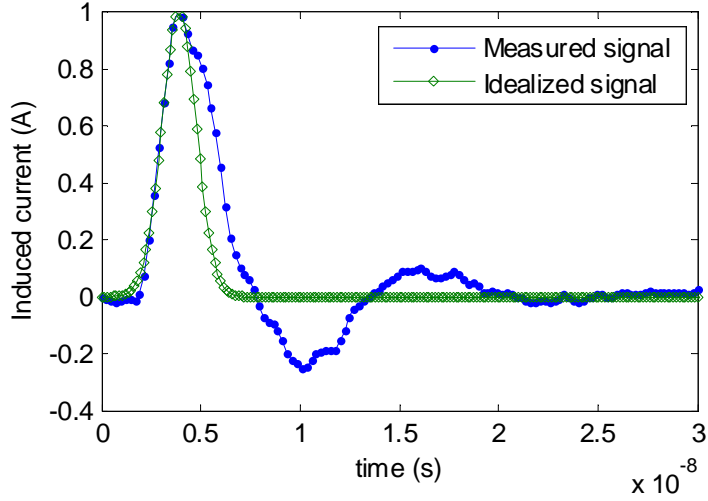

Fig. 2 – Idealized signal used in the examples for clarity, compared with a signal measured in a 0.3 mm gap RPC.

### 3.1. Compensated line

As already mentioned, when the effect of the modes can be neglected, either by the line being sufficiently short or by compensation (making $u = u_0$), the transmission of the injected current to the terminations is determined by $\mathbf{H}(k)$ alone. For the examples in this section the $\mathbf{C}_0$ values were mathematically adjusted for exact compensation while keeping the same $\overline{v}$.

Except for the imaginary capacitive-like term in the main diagonal that depends on the losses, $\mathbf{H}(k)$ is independent of frequency, so there will be no waveform changes but just scaling and coupling to the nearest neighbour (in first order of course). The relative crosstalk is then given by $\eta(k)$, which, for the direct + wave, is just

$$\eta(0) = \frac{1}{1+w} \frac{u+u_0}{2} \tag{35}$$

decreasing with $R_T$ (if $R_T < Z_c$), $u$ and $u_0$.

In Fig. 3 it is plotted both the exact solution (14) and the hypothetical all-strips first-order solution, incorporating (33) into (17) for the parameter set $\mathcal{A}$ in Table 1. This parameter set corresponds to the conditions for which the first-order solution should be most accurate (mathematically compensated, no second-neighbour capacities). Indeed, it can be verified a general very good agreement for the 2 central strips, except for the first reflection of the + wave ($k = 2$) that is suppressed by the $1 - (2k+1)w$ factor in (32), while in the exact solution there is no full suppression. The tentative all-strips solution is accurate for the direct + wave ($k = 0$), as it has been shown in [4] eq. (58), but not for the successive reflections.

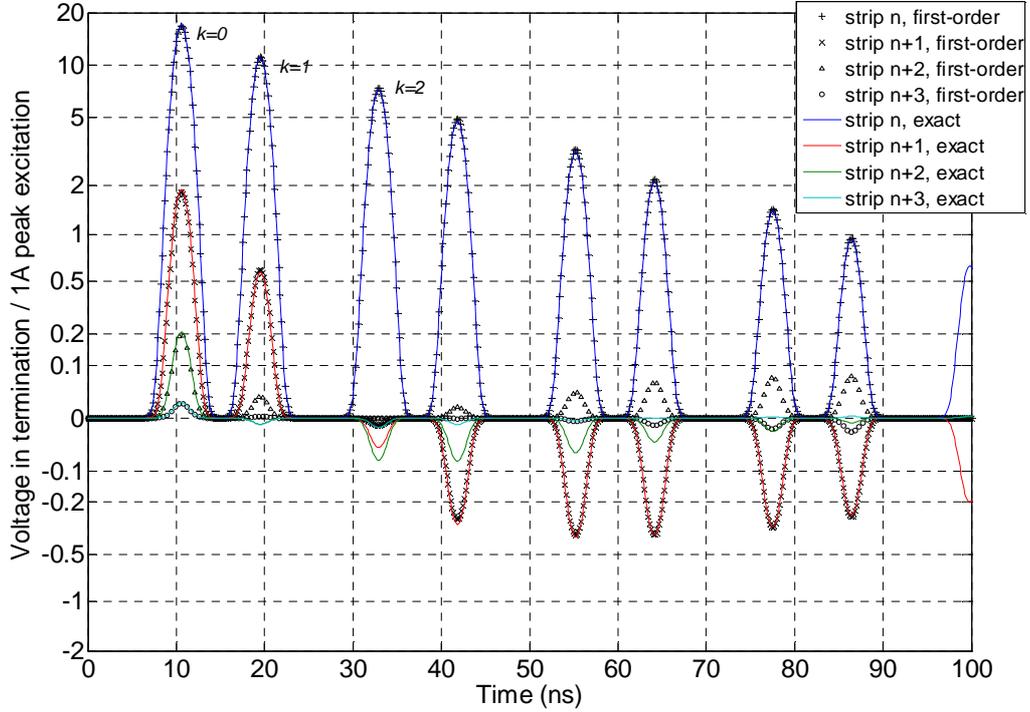

Fig. 3 – Example of the expected signal in the D-end terminations of the central strip and of its immediate neighbours, both for the exact solution (14) and for the tentative all-strips first-order solution, following (33) and (18). Parameter set $\mathcal{A}$. The current in the same terminations can be found by division by $R_T = 100\,\Omega$. Notice in all figures the strongly non-linear vertical scale.

Fig. 4 is similar to Fig. 3, but now using the parameter set $\mathcal{B}$. This includes $2^{nd}$-neighbour capacities ($C_{m2}$) and resistive series ($\mathbf{R}$) losses. The effect of the series losses is to cause a level shift after each impulse with a sign that depends on the sign of the imaginary part of (32). Parallel ($\mathbf{G}$) losses would cause a level shift of opposite sign and may cancel the series losses (Heaviside's cancelation). For comparison, the central strip waveform for the lossless case with all other parameters the same is underlaid, evidencing the signal attenuation owed to the losses. The agreement between the exact and first-order solutions for the 2 central strips remains quite good, the main deviation being that there is a small baseline shift for the neighbour strip that is second-order, so it is not captured by the approximate solution. The exact $2^{nd}$ and $3^{rd}$ neighbour waveforms are now slightly taller than in Fig. 3 owing to the inclusion of the $2^{nd}$ neighbour capacities, but the dominant effect remains the successive nearest-neighbour effect embodied in (33).

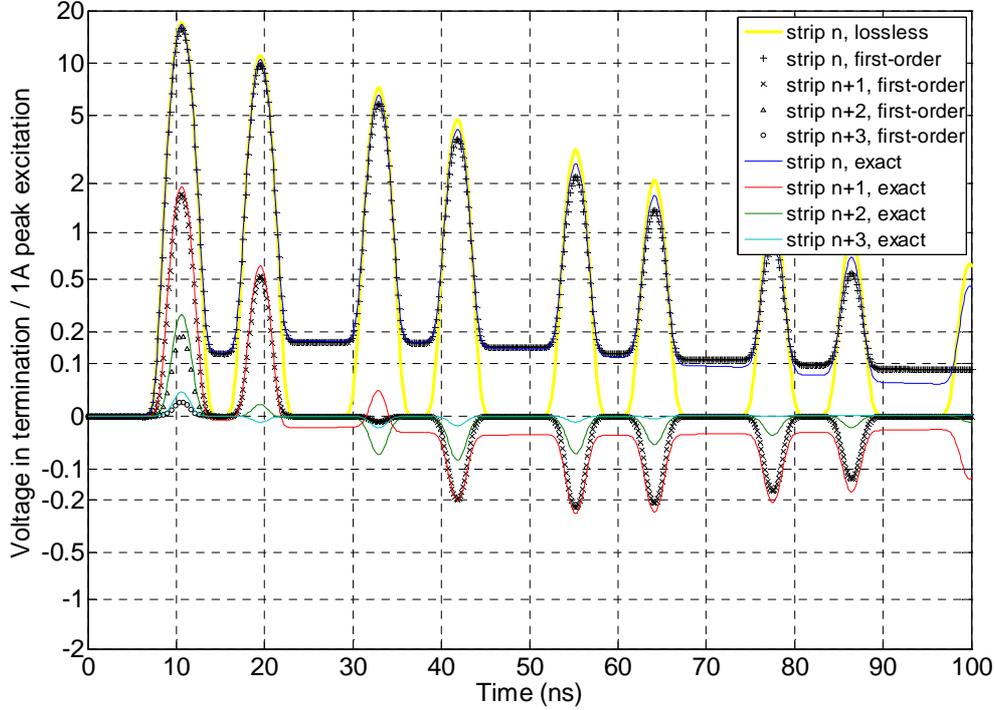

Fig. 4 – Similar to Fig. 3, but now with the parameter set $\mathcal{B}$ (including resistive losses and $2^{nd}$-neighbour capacities). For comparison, the central strip waveform for the lossless case with all other parameters the same is underlaid.

### 3.2. Uncompensated line

As each mode composing the travelling signal propagates at a different velocity and carries a copy of the signal, eventually, given a long enough propagation distance, several instances of the signal would be visibly separated in time. The maximum deviation from the mean propagation time in a single pass is (recall (28))

$$\Delta t = \frac{\Delta v}{\bar{v}^2} D = |u - u_0| \frac{D}{\bar{v}} \tag{36}$$

($D/\bar{v}$ is the average single-pass signal transit time) and gives the scale for the distortion/decoherence of impulses: shorter ones will strongly decohere and much wider ones not so much. Similarly, a slow shaping in the electronics will increase the effective signal width and mitigate the effect, as suggested in [3].

Taking as a model an (analytically convenient) pulse of half width at half maximum $\sigma$, $p(t) = 1/\left(1 + (t/\sigma)^2\right)$, the maximum relative amplitude of a small decoherence $\Delta t$ is, to the first order in $\Delta t/\sigma$,

$$\max\left[\frac{p(t)-p(t+\Delta t)}{p(0)}\right] \approx \frac{3^{3/2}}{8}\frac{\Delta t}{\sigma} \tag{37}$$

This is to be compared with the coupling crosstalk, which, following (35) and in the spirit of (33), should be compared with $\eta^l(0)$ for the $n+l$ strip. Therefore, the relative strength of modal dispersion crosstalk to coupling crosstalk will be approximately

$$\Theta(l) \approx \frac{3^{3/2}}{8\sigma}\frac{D|u-u_0|}{\bar{v}}\left(\frac{1+w}{u+u_0}\right)^l \tag{38}$$

In Fig. 5 it is illustrated the effect of modal decoherence. The conditions are (parameter set $C$) similar to Fig. 4 but waiving mathematical compensation and losses. For comparison the lossless, compensated, case for the central strip is underlaid. The first-order approximation is less accurate in this case, but still reasonable for the central and 1st-neighbour strips. The most striking difference is the considerable increase in crosstalk, compared with the fully compensated case (Fig. 4), for the neighbouring strips, owing to the loss of cancelation between the modes. This is stronger for the 2nd and 3rd neighbours than for the 1st, in consonance with the values of the parameter $\Theta(l)$ shown in Table 1.

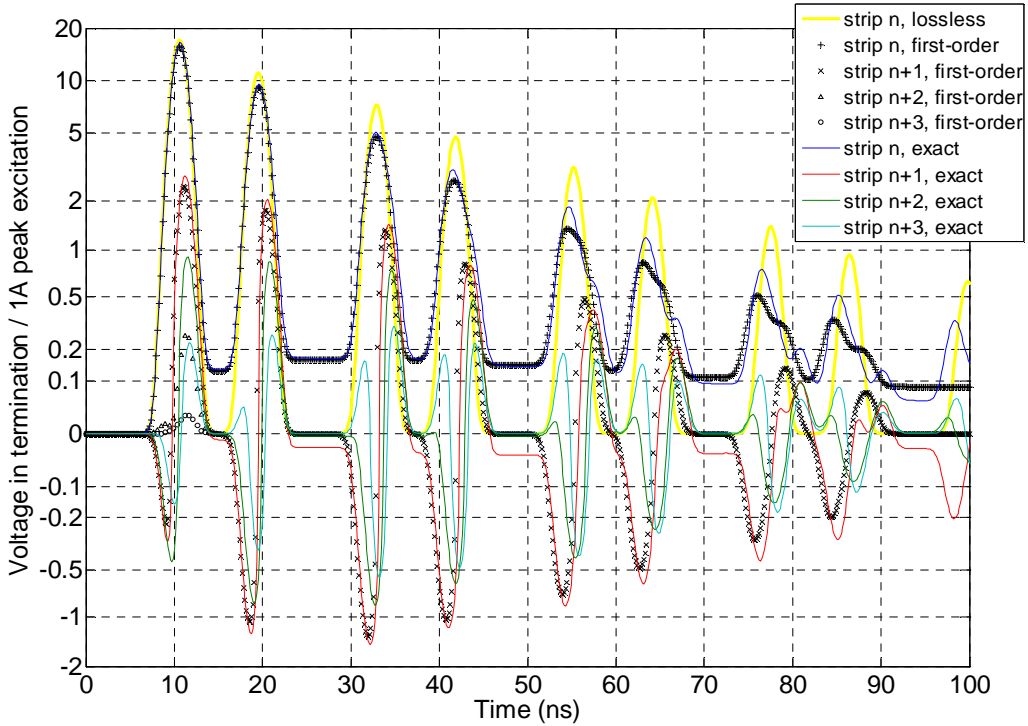

Fig. 5 - Similar to Fig. 4, but now with the parameter set $C$, evidencing the dramatic effect of modal dispersion on crosstalk.

However, it is possible [4] to design well-compensated chambers in stripline structure ("symmetric RPC"), as exemplified in Fig. 6.

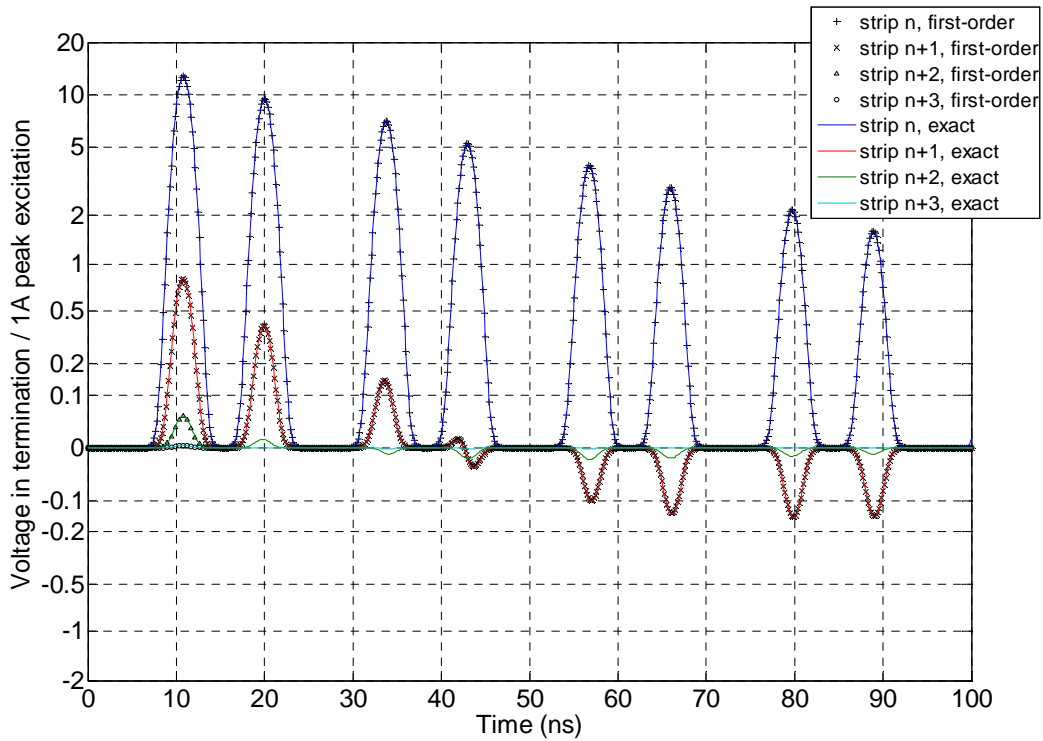

Fig. 6 – Example of a compensated stripline-type RPC [4] from parameter set $\mathcal{D}$.

## 4. Conclusion

The frequency-domain formulation of the multiconductor transmission line (MTL) theory adapted to the case of RPCs (excitation by a current source inside the line itself) allows to treat losses and yields general and compact exact expressions for the signal propagation in RPCs with multiple strip readout electrodes. These expressions can only be evaluated numerically.

In case the interstrip capacities are much smaller than the self-capacity of each strip (weak-coupling), losses are also small and terminations are simple resistors to ground, a first-order development yields simple, fully analytical, expressions valid for all reflections in the central and first-neighbour strips. The modal spectrum can be also calculated, allowing to place qualitative bounds on the signal distortion or excess crosstalk caused by modal dispersion.

Numerical examples confirm that for realistic cases the first-order approximation is enough to describe the two central strips. A simple extension describes well all strips for the direct signal (no reflections) when the chambers are well compensated, that is, modal dispersion is small.

Besides the exponential amplitude reduction of the signal propagating along the chamber, the effect of small losses is to cause a baseline shift that persists after each impulse. Series (resistive) losses and parallel (conductive) losses cause opposite polarity shifts and both effects may cancel (Heaviside's cancelation).

## Acknowledgement

This work was financed by the Portuguese Government through FCT-Foundation for Science and Technology and FEDER/COMPETE under the contract CERN/FP/123605/2011.